\documentstyle[12pt]{article}
\newcommand{\BEQ}{\begin{equation}}
\newcommand{\NEQ}{\end{equation}}

\def\xt{\tilde{x}\,}

\def\Pt{\tilde{P}\,}
\makeatletter
\@addtoreset{equation}{section}
\makeatother

\title{Sufficient Conditions for a Linear Estimator 
to be a  Local Polynomial Regression
}

\author{
Alexander Sidorenko  \/ \/ \/
and  \/ \/ \/ Kurt S. Riedel \\
Courant Institute of Mathematical Sciences \\
New York University \\
New York, New York 10012-1185}

\date{1993 }

\begin{document}

\maketitle

\begin{abstract}
It is shown that any linear estimator
that satisfies the moment conditions up to order $p$
is equivalent to a local polynomial regression of order $p$ with
some non-negative weight function
if and only if the kernel has at most $p$ sign changes.
If the data points are placed symmetrically about the estimation
point, a linear weighting function is equivalent to the standard
quadratic weighting function.
\end{abstract}

\newpage

\section{ Local Polynomial Regression} \label{LPR1}

We consider a linear estimate of a  function or its derivatives
given a sequence of measurements, $\{y_i, i= 1 \ldots N \}$
at the locations, $\{x_i\}$.
In nonparametric estimation, typical assumptions are: $f(t)$ has $p$
continuous derivatives ($q<p$) and 
$y_i=f(x_i)+\varepsilon_i$,
where the errors, $\varepsilon_i$, are independent random variables
with zero mean and variance equal to $\sigma_i^2$.
These assumptions motivate our work, but are not necessary for our results.

One method to select the coefficients of a linear estimator 
is local polynomial regression (LPR) 
as described in works by
Cleveland (1979), 
Fan and Gijbels (1992),
Fan (1993), Hastie and Loader (1993)). 
Not every weighted linear estimate arises from LPR.
We show that {any linear estimator
that satisfies the moment conditions up to order $p-1$
is equivalent to a local polynomial regression of order $p-1$
if and only if the kernel has at most $p-1$ sign changes}.




Let $\{(x_i,y_i),\ i= 1, \ldots N \}$ be given, where
$N$ is the number of measurements, $x_i$ is the $i$th measurement location
and $y_i$ is the corresponding measured value. 
We consider linear estimators of the $q$th derivative of an unknown function,
$f^{(q)}(t)$, of the form:
\begin{equation}\label{WLE}
\widehat{f^{(q)}}(t)  = \sum_{i=1}^{N}  K_i(t) y_i
\ ,
\end{equation}
where the $K_i(t)$ depend on the design, $\{ x_i \}$, but are independent
of $\{y_i\}$.
For a given value of $t$,
we say the weight coefficients, $\{K_i(t),\ i=1 \ldots N\}$,
are of type $(q,p)$ if
it satisfies the {\em moment conditions}:
\BEQ  
\frac{1}{m!} \sum_{i=1}^{N} (x_i-t)^m  K_i(t)  
=  \delta_{m,q} \ , \ \  m=0,\ldots ,p-1
\ . 
\NEQ


In local polynomial regression, at each  
point, $t$, a set of {\bf nonnegative} weights is specified, $w_i(t)$.
and a low order polynomial is fitted to the
weighted sum of squares.
(The weights are usually scaled as
$w_i(t) = W \left(\frac{x_i-t}{h}\right)$,
where $W$ is a non-negative function on [-1,1]
and $h$ is the bandwidth parameter.)
At point $t$, the local estimate of $f(x)$ is
$\sum_{j=0}^{p-1} a_j(t) x^j$, where $p-1$ the order 
of the polynomial approximation.  
The coefficients, $a_j(t)$, are determined by minimizing
\[
 F(a_0,a_1,\ldots,a_{p-1}) \; = \;
 \sum_{i=1}^N w_i(t) \cdot
 \left(\sum_{j=0}^{p-1} a_j(x_i-t)^j - y_i \right)^2
\ . \]
The resulting estimate of $f^{(q)}(t)$is $q!a_q$.
Since the functional is quadratic and non-negative,
the minimum exists and satisfies
\[
 0 \; = \; \frac{\partial F}{\partial a_k} \; = \;
 \sum_{j=0}^{p-1} \left[ \sum_{i=1}^N (x_i-t)^{k+j} w_i(t) \right] a_j
  \; - \; \sum_{i=1}^N (x_i-t)^k w_i(t) y_i
\]
for $k=0,1,\ldots ,p-1$.
This system of linear equations can be rewritten as
\begin{equation}\label{E59}
 \sum_{j=0}^{p-1} d_{kj}(t) \left( a_j h^j\right) \; = \; m_k(t) \; ,
 \;\;\;\;\;\;\;\; k=0,1,\ldots ,p-1 \; ,
\end{equation}
where
\[
 d_{kj}(t) \; = \; \frac{1}{Nh}
   \sum_{i=1}^N \left(\frac{x_i-t}{h}\right)^{k+j} w_i(t)
 \; , \;\;\;\;\;\;\;\;\;\;\;\;\;\;
 m_k(t) \; = \; \frac{1}{Nh}
   \sum_{i=1}^N \left(\frac{x_i-t}{h}\right)^k w_i(t) y_i
 \; .
\]
In (\ref{E59}), $h$ is used solely to scale the equations for numerical
stability. 

If the number of data points with non-zero weights is at least $p$,
the matrix $[d_{kj}(t)]$ is non-singular.
Let $[\tilde{d}_{jk}(t)]$ be the inverse matrix.
Solving for $a_q h^q$ shows that 
vlocal polynomial regression corresponds to
a  linear estimate (\ref{WLE}) with weighting coefficients, $K_i(t)$:
\begin{equation} \label{E60}
 K_i(t) \; = \;  w_i(t)
                   \left[ \frac{q!}{Nh^{q+1}}
    \sum_{k=0}^{p-1} \tilde{d}_{qk}(t) \left(\frac{x_i-t}{h}\right)^k
                   \right]
 \; .
\end{equation}
Let $\xt$ be a dummy variable and  define
\begin{equation}\label{E61}
 \Pt(\xt;t,\{x_i\}) \; = \;  w_i(t)
                   \left[ \frac{q!}{Nh^{q+1}}
    \sum_{k=0}^{p-1} \tilde{d}_{qk}(t;\{x_i\}) \xt^k 
                   \right]
 \; .
\end{equation}
Here $\Pt(\xt;t,\{x_i\})$ is a polynomial of order $p-1$ in $\xt$ 
given $t$ and $\{x_i\}$.
We name $\Pt(\xt;t,\{x_i\})$  the {\em factor polynomial}.
The $p$ coefficients, $\tilde{d}_{qk}(t;\{x_i\})$ determine the
$N$ values of the linear weights: $K_i(t)=w_i(t)\Pt(\xt_i;t,\{x_i\})$
(see M\"uller (1987)).

Thus
{\em
for a given estimation point $t$ and weights $w_i$,
the local polynomial regression estimator
is equivalent to a kernel estimator whose kernel
is the product of the weights with
a polynomial in $\frac{x_i-t}{h}$ of order $p-1$.
}
The equivalent kernel automatically satisfies
the moment conditions and thus is a kernel of type $(q,p)$.

We say that a discrete function $Q(x_i)$ has a sign change between
$x_j$ and $x_{j+k}$ if $Q(x_j)Q(x_{j+k})<0$
and $Q(x_{j+1})=\ldots =Q(x_{j+k-1})=0$.
The weights, $w_i(t)$, are non-negative,
and the factor polynomial has at most $p-1$ roots.
Therefore, for the given $t$,
the equivalent kernel $K(t,x_i)$ has at most $p-1$ sign changes.
Answering the question: ``which kernel estimators can be represented
as a local polynomial regression?''
we show that the necessary condition is also sufficient.


\noindent {\bf Theorem~1.}
{\em
A linear estimator of type $(q,p)$ is 
generated by
local polynomial regression of degree $p-1$ with non-negative weights
if and only if the kernel has no more than $p-1$ sign changes.
}

It is known (see M\"uller (1985)) that any kernel of type $(q,p)$
has at least $p-2$ sign changes. This implies

\noindent {\bf Corollary.}
{\em
The order of the factor polynomial for a degree ($p-1$) LPR 
is at either $p-1$ or  $p-2$.
}

\section{Equivalence of Linear and Quadratic Weightings}\label{LPR2}

It is known (M\"uller (1987), Fan(1993))
that the optimal interior kernel
of type $(q,p)$,
$p-q\equiv 0\,{\rm mod}\, 2$,
is produced by the scaling weight function, $W(y)=1-y^2$,
in the limit of nearly equi-spaced measurement points 
as $N \rightarrow \infty$.
We show that this choice is not unique.

\noindent {\bf Theorem~2.}
{\em
Let $p-q$ be even.
If data points, $x_i$, 
are symmetric around the estimation point, $t$,
and their weights are chosen as $w_i=W\left(\frac{x_i-t}{h}\right)$,
then each of the functions
$W_1(y)=1-y$, $W_2(y)=1+y$, $W_3(y)=1-y^2$
produces the same estimator.
}

The weights, $W_1(y)$ and $W_2(y)$, assign less weight to the estimation point,
$t$, than to one side of the data. This surprising result is useful
in constructing optimal boundary kernels which depend continuously on the
estimation point.

Because of the optimality in the interior,
the Bartlett-Priestley weighting, $W(y)=1-y^2$,
is used often in the boundary region as well
(Hastie and Loader (1993)).
In a future work, we show that a linear weighting has a lower 
asymptotic MSE than the Bartlett-Priestley weighting
in the boundary region.
Theorem 2 shows that one can switch the weighting function
from $W(y)=1-y^2$ to $W(y)=1-y$ without generating a discontinuity in 
the estimate.


\vspace{8mm}

\noindent
{\Large
{\bf Appendix.\/\/ Proofs} 
}


\vspace{3mm}

{\bf Lemma.}
{\em
Let $K_{1,i}$ and $K_{2,i}$ be kernels of type $(q,p)$
with the same estimation point and the same support
such that $K_{r,i}=W_i Q_r(x_i)\;$, $r=1,2$,
where $W_i\geq 0$ for all data points $x_i$ in the support.
If $Q_1(x)$ and $Q_2(x)$ are polynomials of order $p-1$
then $K_{1,i}=K_{2,i}$ for every data point $x_i$.
}

\noindent
{\bf Proof.}
Since $K_1$ and $K_2$ satisfy the same moment conditions,
their difference is orthogonal to any polynomial $P(x_i)$
of order $p-1:\;$
$
 \sum_i (K_{1,i}-K_{2,i})P(x_i) = 0
 .
$
When we choose
$P(x_i)=Q_1(x_i)-Q_2(x_i)$,
we have
$
 \sum_i W_i (Q_1(x_i)-Q_2(x_i))^2 = 0
 .
$
Since $W_i\geq 0$, it implies
$
 W_i (Q_1(x_i)-Q_2(x_i)) = 0
$
for every $x_i$.

{\bf Proof of Theorem~1.}
Let a kernel $K(x_i)$ have $m\leq p-1$ sign changes.
We enumerate the sign changes: $z_1,z_2,\ldots ,z_m$.
Namely, if the $l$th sign change occurs
at $x_j$ or between $x_j$ and $x_{j+k}$,
we set $z_l=x_j+\varepsilon$ where
$\varepsilon < \min\{ x_2-x_1,x_3-x_2,\ldots ,x_N-x_{N-1}\}$.
Now we define $H(x)=(-1)^s \prod_{l=1}^m (x-z_l),\:$
$W(x_i)=K(x_i)/H(x_i)$.
The function $W(x_i)$ has no sign changes.
We choose $s$ to make all of the values $W(x_i)$ non-negative.
Let $Q$ be the factor polynomial for
the local polynomial regression with the weights $w_i=W(x_i)$.
Since $K=WH$ and $WQ$ are kernels of type $(q,p)$,
and $H,Q$ are polynomials of order $p-1$,
The lemma implies that
$K(x_i)=W(x_i)H(x_i)=W(x_i)Q(x_i)$
for every data point $x_i$.
Thus $K$ is the equivalent kernel for
the local polynomial regression with the weights $w_i$.

{\bf Proof of Theorem~2.}
It is sufficient to check that weightings $W_1(y)=1-y$ and $W_3(y)=1-y^2$
have the same equivalent kernel.
Let $Q_1(y)$ and $Q_3(y)$ be their respective factor polynomials.
Since $Q_3$ is a polynomial of order $p-1$,
then $W_3Q_3$ is a polynomial of order $p+1$.
Since $W_3$ is even
and the placement of data points is symmetric,
the equivalent kernel, $W_3Q_3$, is an even function
(if $q$ is even) or an odd function (if $q$ is odd).
The difference $p-q$ is even, thus $W_3Q_3$
can not have term $y^{p+1}$.
Therefore, $W_3Q_3$ is a polynomial of order $p$,
and the true order of $Q_3$ is at most $p-2$.
Now we notice that $W_3(y)Q_3(y)=W_1(y)\left[ (1+y)Q_3(y)\right]$.
Both $(1+y)Q_3(y)$ and $Q_1(y)$ are polynomials of order $p-1$.
Thus the lemma implies that $W_3(y)Q_3(y)=W_1(y)Q_1(y)$
when $y=\frac{x_i-t}{h}$.



\noindent {\bf References}

Cleveland, W. S. (1979).
Robust locally weighted regression and smoothimg scatterplots.
{\em J. Amer.\ Statist.\ Assoc.} {\bf 74} 829-836.

Fan, J. and Gijbels, I. (1992).
Variable bandwidth and local linear regression smoothers.
{\em Ann.\ Stat.} {\bf 20} 2008-2036.

Fan, J. (1993).
Local linear regression smoothers and their minimax efficiencies.
{\em Ann.\ Stat.} {\bf 21} 196-216.



Hastie, T. and Loader, C. (1993).
Local regression: automatic kernel carpentry.
{\em Statistical Science} {\bf 8} 120-143.



M\"uller, H. G. (1985).
On the number of sign changes of a real function.
{\em Periodica Mathematica Hungarica} {\bf 16} 209-213.

M\"uller, H. G. (1987).
Weighted local regression and kernel methods
for nonparametric curve fitting.
{\em J. Amer.\ Statist.\ Assoc.} {\bf 82} 231-238.



\end{document}